\documentstyle[onecolumn]{mn}
\def\lsim{~\rlap{$<$}{\lower 1.0ex\hbox{$\sim$}}}
\def\bsim{~\rlap{$>$}{\lower 1.0ex\hbox{$\sim$}}}

\def\dd{{\rm d}}
\def\ln{{\rm ln}}
\def\pa{\partial}

\def\pmb#1{\setbox0=\hbox{#1}%
\kern-.025em\copy0\kern-\wd0
\kern.05em\copy0\kern-\wd0
\kern-.025em\raise.0433em\box0}

\def\vV{\pmb{$V$}}
\def\vv{\pmb{$V$}}
\def\valpha{\pmb{$\theta$}}

\def\vtheta{\pmb{$\theta$}}

\def\vg{\pmb{$g$}}

\def\vx{\pmb{$x$}}
\def\vs{{\pmb{$s$}}}

\def\vC{\pmb{$C$}}

\def\da {{\dot a}}

\def\vnabla{\pmb{$\nabla$}}

\def\xpar{{\vx^\parallel}}
\def\xper{{\vx^\perp}}
\def\cpar{{\vC^\parallel}}
\def\cper{{\vC^\perp}}

\begin{document}
\title[On the least action principle]
{On the least action principle in cosmology}
\author[Nusser \& Branchini]{Adi Nusser$^1$ and Enzo Branchini$^2$
\\
$^1$ Physics department,
Technion, Haifa 32000, Israel
\\
$^2$ Kapteyn Institute, University of Groningen, Landleven 12, 9700 AV, 
Groningen, The Netherlands}
\maketitle

\begin{abstract}
Given the present distribution of mass tracing objects in an
expanding universe, we develop and test a fast method for recovering
their past orbits using the least action principle.  In this method,
termed FAM for Fast Action Minimization, the orbits are expanded
in a set of orthogonal time-base functions satisfying the appropriate
boundary conditions at the initial and final times.  The conjugate
gradient method is applied to locate the extremum of the action in the
space of the expansion coefficients of the orbits.  The TREECODE
gravity solver routine is used for computing the gravitational fields
appearing in the action and its gradients.  The time integration of
the Lagrangian is done using Gaussian quadratures.  FAM allows us to
increase the number of galaxies used in previous
numerical action principle
implementations by more than one order of magnitude. For example,
orbits for the $\sim 15,000$ IRAS PSC$z$ galaxies can be recovered in
$\sim 12,000$ CPU seconds on a 400MHz DEC-Alpha machine.  FAM can
recover the present peculiar velocities of particles and the initial
fluctuations field.  It successfully recovers the flow field down to
clusters scales, where deviations of the flow from the Zel'dovich
solution are significant.  We also show how to recover orbits from the
present distribution of objects as in redshift space by direct
minimization of a modified action, without iterating the solution
between real and redshift spaces.

\end{abstract}
\section {Introduction}

In the standard cosmological paradigm, the present distribution of
galaxies and their peculiar motions are the result of gravitational
amplification of tiny initial density fluctuations.  Accordingly, the
mildly non-linear large scale structure observed today contains
valuable information on the initial fluctuations.  Given either the
present galaxy distribution or the peculiar velocity field one can
recover the growing mode of the initial density field 
(Peebles 1989, Weinberg 1992,
Nusser \& Dekel 1992, Gramman 1993, Giavalisco et al
1993, Croft \& Gaztanaga 1998, Narayanan \& Weinberg 1998). Methods for recovering the initial
fluctuations can be very rewarding as one can directly address 
 statistical properties of these fluctuations (Nusser \& Dekel
1993). As has been shown by Nusser \& Dekel (1992,1993) the initial
density field recovered from peculiar velocity fields has a strong
dependence on the value of the density parameter, $\Omega$.  On the
other hand, a recovery from the galaxy distribution depends very weakly
on $\Omega$. Therefore by matching the statistical properties from the
two recovered initial density fields one could provide an estimate for
$\Omega$.  Gravitational instability theory, also provides tight
relations between the present density and peculiar velocity fields
(e.g., Nusser et al 1991). These relations have been an important tool
in large scale structure studies, in particular for model independent 
estimates of the cosmological parameters. For example, 
a comparison of the measured peculiar
velocities with those predicted from galaxy redshift surveys 
can yield $\Omega$.
(Davis, Nusser \& Willick 1996, Willick et al 1996, da Costa et al 1998,
Sigad et al 1998, Branchini et al 1999).
These comparisons are done under an assumed form for the
biasing relation between the mass and galaxy distribution. So deviations 
from the theoretical velocity-density relations may serve as an indication   
to the way galaxies trace the mass and hence to the interplay between 
galaxy formation and the large scale environment.

Most
of the promising methods for
recovering the initial growing mode and for relating the present 
peculiar velocity and density fields, are based on the least action
principle (LAP).  Peebles (1989) (hereafter P89) has pointed out that the
equations of motion can be derived from the stationary variations of
the action with respect to 
orbits subject to  fixed final positions and  vanishing initial 
peculiar velocities.
P89 proposed
minimizing the action with respect to the coefficients of an expansion
of the orbits in terms of time-dependent functions satisfying the
appropriate boundary conditions. Methods based on LAP 
are very powerful as the  
true orbits are recovered to any accuracy depending
on the number of functions used in the expansion, at least in the 
laminar flow regime. They also provide simultaneous estimates of the 
present peculiar velocities and the initial fluctuations from 
a given distribution of galaxies.

So far, the action principle has been applied to study the dynamics of
the Local Group of galaxies (P89, Peebles 1991, 1994, Dunn \& Laflamme
1993, Branchini \& Carlberg 1994) and to recover the peculiar velocity
from the distribution of about 1100 galaxies within a redshift of
$3000 \ \rm km/s$ (Shaya, Peebles \& Tully 1995).  The application
of LAP to larger data sets have been hindered by the heavy
computational burden needed in current methods for minimizing the
action.  The LAP is the only technique with which one can probe the
nonlinear behavior to any accuracy as opposed to Zel'dovich type
approximations which break down near large mass concentrations like
clusters of galaxies. Therefore, a fast method for implementing LAP
can be rewarding. Such a method would enable us to apply LAP to large
data sets like the various IRAS galaxy redshift surveys (1.2 Jy,
Fisher et al 1995a, and PSC$z$, Saunders 1996) and portions of the
Sloan Digital Sky Survey (Gunn \& Knapp 1993).  In contrast to
previous numerical implementations 
of LAP, which use direct summation over inter-particle
forces, we propose here an efficient method based on the TREECODE to
compute gravitational forces and the potential energy.  Also, our
method expands the orbits in orthogonal time base functions and
perform the time integration using the Gaussian quadratures method.
All this increases the speed of the calculation by more than one order of
magnitude over any previous implementation of LAP. We refer to our
method by the acronym FAM for Fast Action Minimization.

The outline of the paper is as follows. In section 2 we review the
least action principle and describe FAM.
In section 3 we demonstrate the robustness of and
show tests of the recovered peculiar velocities. In section 4 we
describe extensions of the FAM to flux limited surveys, application
from redshift surveys and biased distribution of galaxies.  We
conclude in section 5 with a discussion of our results and possible
applications of FAM.

\section{the boundary value problem and the least action principle}

We follow the standard notation in which $a(t) $ is the scale factor,
$H(t)=\da /a$ is the time-dependent Hubble factor,
$\Omega=\rho_b/\rho_c$ is the ratio of the background density,
$\rho_b$, of the universe to the critical density, $\rho_c=3H^2/8\pi
G$.

Assume that the underlying mass density field in a spherical volume
$V$ is sampled, in an unbiased way, by a discrete distribution of $N$
galaxies (particles). In this sampling, if the average mass density
in any cell of volume $\delta\! V\!<\!<V$ is $\rho_{_{\delta\! V}}$
then the number of particles in that cell is drawn from a Poisson
distribution with mean ${\bar n }(\rho_{_{\delta\! V}}/{
\rho_b})\delta\!V$, where $\bar n=N/V$ is the mean number density over
the large volume $V$ and we have assumed that the average mass density in 
 $V$
is $\rho_b$. Instead of using the usual time variable, $t$,
we describe the evolution of the system in terms of the linear growing
mode, $D(t)$, of density perturbations (e.g., Peebles 1980).  Let $
\vx_i$ denote the comoving coordinate of the $i^{\rm th}$ particle and
$\vtheta_i=\dd \vx_i/\dd D$ its velocity with respect to the time
variable $D$.  Neglecting interactions between matter interior and
exterior to the volume $V$, system of particles obeys the the
following Euler equations
\begin{equation}
\frac{\dd \valpha_i}{\dd D} +\frac{3}{2}\frac{1}{D}\vtheta_i =
\frac{3}{2}\frac{1}{D^2}\frac{\Omega}{f^2(\Omega)}\vg(\vx_i) , 
\label{euler}
\end{equation}
where $f(\Omega)=\dd \ln D /\dd \ln a\approx \Omega^{0.6}$ is the
linear growth factor (e.g., Peebles 1980) and $\vg$ represents the
peculiar gravitational force field per unit mass. These equations are
supplemented by the Poisson equation
\begin{equation} \vnabla \cdot
\vg(\vx)= -\delta(\vx) \, ,
\label{poisson}
\end{equation}
which relates the divergence of $\vg$ to the mass density
contrast  $\delta(\vx)=\rho(\vx)/\rho_b-1$  
at any point $\vx$ in comoving coordinate space. 
We can approximate $\delta$ from the discrete particle distribution by
\begin{equation}
\delta(\vx)= \frac{1}{V} \sum_{i=1}^N\delta^{\rm D}(\vx -\vx_i)-1 \, ,
\end{equation}
where $\delta^{\rm D}$ is the Dirac delta function with unit integral 
over the volume $V$. Therefore, the field $\vg$ is
given by
\begin{equation}
\vg(\vx)=-\frac{1}{4\pi \bar  n }\sum_i \frac{\vx-\vx_i} {|\vx-\vx_i|^3} 
+\frac{1}{3}\vx . \label{pot}
\end{equation}
Equations (\ref{euler}) and (\ref{pot}) constitute the equations of
motion governing the evolution of the system of particles. We do not
include the continuity equation in the equations of motion. The
continuity equation is a constraint equation obeyed automatically as
the particles move according to (\ref{euler}) and (\ref{pot}).  The
equations of motion involve second order time derivatives. Solving
these equations can be seen either as an initial value problem or as a
boundary value problem (Giavalisco et. al. 1993).  Numerical N-body
codes (e.g., Hockney \& Eastwood 1981) are the usual tool for solving
the relevant initial value problem where the positions and velocities
of particles are specified at a given time.  But, recovering the
orbits of particles given their present positions is a boundary value
problem where the solution must yield a homogeneous distribution of
particles at the initial time $D=0$.  P89 stated the cosmological
boundary value problem in the context of the least action principle.
P89 also suggested an approximation to the orbits by minimization of
the action with respect to a particular choice trial
functions.  In our notation, the action of the system of particles is
\begin{equation}
{\mathrm S}=\int_0^{1} \dd D   \sum_i
\left\{\frac{1}{2}  D^{3/2} \vtheta_i^2
+\frac{3}{2}\frac{1}{D^{1/2}}\frac{\Omega}{f^2(\Omega)} 
\left[  \frac{1}{4\pi \bar n}
\sum_{j<i}\frac{1}{|\vx_i-\vx_j|}
+\frac{ \vx_i^2}{6}  \right]   \right\} \, , 
\label{lapD}
\end{equation}
where we have arbitrarily set $D=1$ at the present time. Following  P89
we  minimize the action with respect to the coefficients
of an expansion of the orbits by means of time
dependent base functions $\{q_n(D),n=1\cdots n_{max}\}$. We
write the position of each particle  $\vx_i(D)$ for $D<1$
as 
\begin{equation}
\vx_i(D)=\vx_{i,0}+\sum_{n=1}^{n_{max}} q_n(D) \vC_{i,n}  \, ,
\label{expand} 
\end{equation}
where $\vx_{i,0}$ is the position of the particle at $D=1$ and the 
vectors $\vC_{i,n}$ 
are the expansion coefficients with respect to which the action is to be 
minimized. Since $\vx_i(D=1)=\vx_{i,0}$, we choose the functions  $q_n(D)$
such that  $q_n(D=1)=0$.
By taking derivatives of (\ref{expand}) with respect to $D$,
we find that the velocity, $\vtheta_i$, is
\begin{equation}
\vtheta_i(D)=\sum_{n=1}^{n_{max}} p_n(D) \vC_{i,n}  \, ,
\label{expandv} 
\end{equation}
where $p_n=\dd q_n /\dd D$.
Stationary  variations of the action with respect to $\vC_i$  yield the
following set of equations
\begin{equation}
\frac{\pa {\mathrm S}}{\pa \vC_{i,n}}=
\int_0^1 \dd  D \left[ D^{3/2} p_n \vtheta_i +\frac{3}{2} 
\frac{q_n} {D^{1/2}} \frac{\Omega}{f^2(\Omega)}\vg_i \right] = 0 \, .
\label{ceqn}
\end{equation}
If we integrate by parts the term involving the velocity in the previous 
equation, we arrive at
\begin{eqnarray}
\left(D^{3/2} q_n \vtheta_{i}\right)_{D=1} &-&
\lim_{D\rightarrow 0}\left(D^{3/2} q_n \vtheta_i \right)
- \nonumber \\ 
&&\int_0^1 \dd  D D^{3/2} q_n(D) \left[  
\frac{\dd \valpha_i}{\dd D} +\frac{3}{2}\frac{1}{D}\vtheta_i
-\frac{3}{2}\frac{1}{D^2} \frac{\Omega}{f^2(\Omega)}\vg_i \right] = 0 \, . 
\nonumber \\
\label{ceqnparts}
\end{eqnarray}
Without the boundary terms on the left, 
these equations are equivalent
to the equations of motion averaged over time with weight functions 
$D^{3/2}q_n$.
The boundary terms are individually eliminated by the 
imposing the following two constraints on the time-base 
functions (Peebles 1989),
\begin{eqnarray}
q_n(D=1)=0 \quad & {\rm and} & 
\qquad \lim_{D\rightarrow 0} D^{3/2} q_n(D)\vtheta(D)=0 \, .
\label{boundary}
\end{eqnarray}

\subsection{The homogeneity condition and the time-base functions}
We will work 
directly with the functions $p_n=\dd q_n/\dd D$ rather than $q_n$.
The first constraint in (\ref{boundary}) means that the positions of
particles at $D=1$ remain unchanged when varying $\vC_{i,n}$.  The
second constraint is very flexible, it  merely implies that 
$\dd \ln p_n(D)/\dd \ln D >-5/4$ for
$D\ll 1$.  However the functions $p_n$ must  
lead to homogeneous initial particle distribution.
A sufficient condition for the homogeneity is that
the time dependence of the particle velocities  near $D=0$ matches
that of the linear velocity growing mode which, with respect to the time $D$,
is a constant.  Orbits with velocities with initial time dependence
like that of the decaying mode ($\propto D^{-5/2}$), do not
necessarily lead to homogeneous initial distributions, although the
decaying mode is derived under the assumption of small perturbations.
Therefore, initial homogeneity implies that one of the
$p_n$ must be a constant and the rest increasing functions of $D$.
There are many functions satisfying our boundary conditions.  Here we
choose  $p_n$ to be linear combinations of  $1,D, D^2\cdots
D^{n_{max}}$ (Giavalisco et. al. 1993)
which  satisfy  the following orthonormality
condition
\begin{equation}
\int_0^1 \dd D D^{3/2} p_n(D) p_m(D)=\delta^{\rm K}_{m,n} \label{ortho}
\end{equation}
where $\delta^{\rm K}$ is the Kronecker delta function.
Orthonormality  will prove useful in the 
numerical minimization of the action using the conjugate gradient method.
The functions $p_n$ can be constructed using the Gramm-Schmidt algorithm,
however, in this case they can be derived  from the expression
\begin{equation}
p_n(D)=A_n \frac{1}{D^{3/2} }\frac{\dd^n}{\dd D^n}\left[
D^{3/2} D^n (1-D)^n \right] , \label{hn}
\end{equation}
where $A_n$ are normalizing constant.
With the orthonormality condition (\ref{ortho}) 
the expression  for the action gradients 
$\pa {\mathrm S}/\pa C_i^\alpha$ (\ref{ceqn}) becomes 
\begin{equation}
\frac{\pa {\mathrm S}}{\pa \vC_{i,n}}=
\vC_{i,n} +
\frac{3}{2}\!\int_0^1\! \frac{q_n}{D^{1/2}}
\frac{\Omega}{f^2}\,\vg_i(D)\, \dd D\, .
\label{ceqn:ortho}
\end{equation}

\subsection{The numerical action minimization}
\label{FFF}
Given the orbit expansion (\ref{expand}) and the base-functions (\ref{hn}), 
the problem of recovering the orbits reduces to finding the value of
the coefficients $\vC_{i,n}$ where the action (\ref{lapD}) has a
minimum.  Our method for minimizing the action, FAM, is based  on
the conjugate gradient method (CGM) (e.g., Press et. al. 1992).  An
efficient implementation of CGM calls for a fast way of
computing the action and its gradients with respect to $\vC$.  Most of
the computational cost comes from the potential energy term in the
action and the gravitational force field, $ \vg$, in the gradients
(\ref{ceqn:ortho}).  These quantities involve summation over pairs and
they have to be computed various times, in each step taken by CGM, for
an accurate estimate of their integrals.  We can achieve a significant
improvement over previous schemes for minimizing the action if instead
of computing the gravitational forces by direct summation, we use the
TREECODE technique (e.g., Bouchet \& Hernquist 1988).  Although any of
the fast techniques like the particle-mesh (PM) or
particle-particle particle-mesh (P$^3$M) (Hockney \& Eastwood 1981) or the
adaptive P$^3$M (Couchman 1991) can be used, the TREECODE is particularly
suitable for our purposes since it can readily be implemented for
particle distributions in a spherical region as in whole
sky galaxy catalogs.  We reduce even further the required number of
gravitational field calculation by performing the time integration
 in the expression for the action and its gradients
using the Gaussian quadrature scheme with $D^{-1/2}$ weights
(Giavalisco et al 1993).

The boundary value problem is almost certain to have more than one
solution (P89, Giavalisco et. al. 1993).  Part of the reason for this
is that one of the boundary conditions only prescribes time dependence
of the velocities near the initial time $D=0$ and does not specify
their amplitude.  This is not sufficient for a unique solution, in the
presence of orbit mixing regions.  So the action can have many minima
corresponding to different solutions of the time averaged equations of
motion.  Therefore, we expect the minimum found by CGM to depend on
the initial guess.  However, 
our purpose is to recover motions on large scales which means that
out of all minima we would like to find
the one corresponding to orbits which do not deviate significantly
from the Zel'dovich straight line approximation.
A reasonable choice for the initial guess
is then: $\vC_{i,n}=0$ for $n>1$,
and $\vC_{i,1}$ obtained from
(\ref{ceqn:ortho}) by substituting   $\vg_i(D)=D\vg_{i,0}$ where 
$\vg_{i,0}$ is the gravitational force 
field at $D=1$.
This means that the initial guess for the orbit of  each particle 
is a motion in straight line
with  velocity  given by  the gravitational
force field according to  linear theory.
\subsection{The dependence on $H_0$ and $\Omega$}

We would like to clarify here the dependence of the equations of
motion and the recovered orbits on the present value of Hubble factor,
$H_0$, and the density parameter, $\Omega$.  Since $D$ is a
dimensionless variable, the spatial coordinate $\vx$ and the velocity
$\vtheta=\dd \vx/ \dd D$ have the same units, e.g., $\rm Mpc$. We
could also work in $\rm km/s$ if we define $H_0 \vx$ to be the
spatial coordinate.  Neither the action nor its gradients
(\ref{ceqn:ortho}) involve the Hubble factor, so the recovered orbits are
completely independent of the units with which we choose to express
the orbits.  

The equations of motion expressed in terms of the time variable, $D$,
are almost independent of $\Omega$ and the cosmological constant
(e.g., Weinberg \& Gunn 1990, Nusser \& Colberg 1997, Mancinelli \&
Yahil 1995, Gramman 1993).  This is because of the weak dependence of
$\Omega/f^2\approx \Omega^{-0.2}$ on the cosmological parameters
in  \ref{euler}.
However, as we shall see in subsection 4.2, the orbits recovered from
the distribution of galaxies in redshift space rather in real space,
will have a non-negligible dependence on $\Omega$.  One could then
obtain orbits in a $\Omega \ne 1$ universe by appropriate scaling
those recovered assuming flat universe.

\section{testing FAM with an N-body simulation}

\begin{figure}
\vspace{13truecm}
{}
\caption{Recovered orbits from the FAM. Filled dots show 
present time positions for a random selection of N-body particles
contained within a slice of  thickens 10 $h^{-1}$ Mpc.
The solid lines represent their projected orbits.
The upper plot is an enlargement of the central region 
in the lower plot.} 
\label{orb}
\end{figure}

FAM involves a number of parameters. These include the force softening
scale in the TREECODE gravity solver, the number of time-base  functions,
 the convergence tolerance parameter in CGM,
 (see Press et. al. 1992 for details).  In addition, we have the 
initial guess  $\vC_{i,n}$ required by CGM. 
We resort to an N-body simulation 
to demonstrate the robustness of the method to these
parameters and  the initial guess.
We use a high resolution simulation (Cole et. al. 1998) of cold dark
matter in a flat universe with a cosmological constant.  The matter
density parameter at the final output of the simulation is
$\Omega_0=0.3$ The simulation contained
192$^3$ particles in a periodic cube of side 345.6 $h^{-1}$ Mpc.  The
simulation is normalized such that at the final output the linearly
extrapolated $rms$ value of the density fluctuations in a sphere of
$8{\rm h^{-1} Mpc}$ is $\sigma_8=1.13$.
The simulation was
run using a modified version of Couchman (1991) AP$^3$M N-body code
with a force softening parameter of 0.27 $h^{-1}$ Mpc.  We test FAM using
the distribution of $1.5(\cdot 10^4)$ particles selected at random from a
spherical region of radius 80 $h^{-1}$ Mpc in the simulation. 
We will refer to a standard
FAM recovery as the one in which the orbits are expanded 
in six time-base functions given by
(\ref{hn}),  
the  tolerance parameter in CGM is $10^{-4}$, and 
the initial guess for $\vC_{i,n}$ is given from linear theory, as described
at the end of subsection (\ref{FFF}).
The total number of free coefficients $\vC_{i,n}$ in standard FAM
is  $3\times 1.5(\cdot 10^4) \times 6=2.7(\cdot 10^5)$,
including the 3 spatial components. 

We will focus on the performance of FAM at recovering the particle
velocities at the final output of the simulation.  However, it is
instructive to visually examine the recovered orbits.
The solid lines in figure 1 are two dimensional streamlines of particles
contained, at the final time, in a slice of thickness of $10h^{-1}$
Mpc.  The dots indicate the present positions of the particles.  The
streamlines shown in the plot correspond to orbits recovered with
standard FAM and a force softening parameter of $0.5h^{-1}{\rm Mpc}$.
The deviations of the streamlines from straight lines are
significant, especially in high density regions. These deviations are
an indication of the failure of the Zel'dovich approximation. 
 
We first assess the robustness of the recovered velocities against
changes in the initial guess for $\vC_{i,n}$.  To do that we have
compared the solution of standard FAM with the solution obtained using
$\vC_{i,n}=0$ as the initial guess.  The three panels on the left hand
side in figure 2 compare between the FAM recovered velocities in the
two cases\footnote{ In this and the following figures we plot comoving
peculiar velocities $\vV= \dd \vx/ \dd t=H_0 f(\Omega_0) \vtheta$
measured in units of $\rm km/s$}.  The three panels show,
respectively, results for three values of the force softening
parameter, as indicated in each panel.  Changing the initial guess did
not introduce any systematic differences in the recovered velocities
for all three values of the softening parameter.  The scatter,
although not negligible,  is small compared to the scatter between the
recovered  true velocities (see figure 4 below).  The right
hand side of figure 2 compares the standard FAM  velocities
with those recovered with a convergence tolerance parameter of 10$^{-5}$.
The correlation between the two velocities is very tight and no
systematic differences are detected.

\begin{figure}
\vspace{13truecm}
{}
\caption{Robustness tests of FAM. {\em To the left}: each panel shows
 velocities (one component) '
recovered with $\vC_{i,n}=0$ as the initial guess $vs$
those recovered with standard FAM.  The panels correspond,
respectively, to different values for the force softening parameter.
{\em To the right}: each panel compares velocities recovered with two
values of the convergence tolerance parameter as indicated at the top.}
\label{vvrob}
\end{figure}  

Having established the robustness of FAM we now proceed to check 
how well
it reproduces true velocities of particles in the simulation.  In
figure 3 we show the scatter plots of recovered $vs$ true velocities
of randomly chosen particles.  In addition to velocities recovered
with standard FAM (left column) we show results from the Zel'dovich
approximation (middle column) and linear theory (right column).
Velocities in the Zel'dovich approximation were obtained by running
FAM with $n_{max}=1$, i.e,  with straight line orbits of the form
$\vx_i(D)=\vx_{0,i}+D\vC_{i,1}$. 
The Zel'dovich velocities 
should coincide with those which would have been recovered 
by the PIZA method of Croft \& Gaztanaga (1998). The linear theory 
velocities are simply $H_0 f(\Omega_0) \vg_{0,i}$ where $\vg_{0,i}$ is 
the gravitational force field obtained from the particle distribution
at the final time.
In all cases the  softening 
parameter is $0.5h^{-1}{\rm Mpc}$ in the computation 
of the gravity field.
The recovered and true velocities in each panel
 with a top-hat window of the same length.
The three rows show results for three smoothing lengths, as 
indicated in the figure.
The top-hat smoothing replaces the velocity of each particle
by the mean velocity of its neighbors within a distance equal to the 
smoothing width.
The parameters of the best linear fit are shown on the top left corner of 
each plot.
Linear theory performs very poorly, even when smoothing on scales as
large as 5 $h^{-1}$.  Evidently, the Zel'dovich approximation is a
significant improvement over linear theory.  Yet, 
a close visual inspection reveals a systematic bias 
in  the Zel'dovich velocities. This is confirmed quantitatively by
the parameters of the linear regression.
The FAM velocities 
are almost unbiased even for the smallest smoothing width.
Note also the small scatter of  150 - 250 $\rm km/s$  between 
the FAM and true velocities.

\begin{figure}
\vspace{13truecm}
{}
\caption{Recovered {\em vs} true velocities for random particles in the 
simulation. Left, middle and right columns correspond, respectively,
to recovery from, standard FAM, Zel'dovich approximation, 
and linear  theory. 
Shown are velocities smoothed with a top-hat window of  
radius  5 (top row),
3  (middle), and 1 $h^{-1}$ Mpc (bottom).
Displayed in each panel, the parameters of 
the linear regression of recovered on true velocities.
The $45^\circ$ solid lines are drawn to guide the eye.}
\label{vsm}
\end{figure} 

The superiority of  FAM over the Zel'dovich approximation  can 
be further appreciated from figure 4 where we plot
the  unsmoothed velocities. For both in FAM and Zel'dovich we 
we  show results with force softening of 0.25 $h^{-1}$ Mpc 
(top panels) and 3 $h^{-1}$ Mpc (bottom panels) 
 in the TREECODE.  The lower softening  matches the 
the one used for the force calculation in the original  N--body simulation.
The Zel'dovich solution fails  
to  reproduce nonlinear motions for both values of the 
softening parameter. It typically underestimates the
true velocities.
FAM, however, provides unbiased estimates of the 
true velocities if 
the force softening is close to the one of the N-body simulation.
FAM cannot model accurately the structure of the orbits
on scales that the force softening length. This is the cause of the 
bias in FAM velocities recovered with the high value of the softening
parameter (bottom right panel).
The  random errors in the unsmoothed FAM velocities  are large 
($\sim \rm 500 km/s$) but they go down  by a factor of 2 when smoothing 
on a scale of 1 $h^{-1}$ Mpc (see bottom left panel in figure 3).

\begin{figure}
\vspace{13truecm}
{}
\caption{FAM {\em vs.} Zel'dovich. Plotted are unsmoothed velocities. 
{\em To the right}:  Zel'dovich {\em vs.} True. {\em To the left}:  
LAP {\em vs.} true.
Shown are velocities recovered with  softening parameters of
  0.25 $h^{-1}$ Mpc (top plots) and
3 $h^{-1}$ Mpc (bottom). The parameters of the linear fit
are shown in each panel}
\label{vvun}
\end{figure}

\section{extensions of FAM}
So far we had in mind an ideal situation in which we 
have a perfect volume limited unbiased distribution
of galaxies in real space. All sky surveys like the IRAS
sample, provide us with the redshift coordinates of galaxies. They are 
typically flux limited so that the observed number density of galaxies is 
a decreasing function of distance. Galaxies may also be biased tracers of the
underlying mass density field.
We will now outline  how FAM can  deal with 
these issues.

\subsection{Selection effects and shot-noise}
Suppose that a flux limited catalog is obtained from a parent 
volume limited catalog.
If the observed number density in real space in a flux limited catalog
is $n_0(\vx)$, then the corresponding
 number density in the volume limited catalog is $n_0(\vx)/\phi(|\vx|)$,
$\phi$ is the selection function. The gravitational field $\vg$ can
then be approximated by
\begin{equation}
\vg(\vx)=-\frac{1}{4\pi   n_1 }\sum_i \frac{1}{\phi_i} \frac{\vx-\vx_i} {|\vx-\vx_i|^3} 
+\frac{1}{3}\vx , \label{gforce}
\end{equation}
where $\phi_i=\phi(\vx_{i,0})$ and $n_1$ is an estimate of the number
density in the volume limited catalog, for example, $n_1\approx \sum_i
\phi^{-1}_i /V$ where the sum is over galaxies within a spherical
region of volume $V$ around the observer.  The potential energy term
in the expression for action is also modified accordingly.  So for
flux limited surveys the gravitational potential and force fields are
computed by the TREECODE technique with each particle having a mass
proportional to the inverse of the selection function at its final
position. We now examine the error in the recovered velocities and
positions of particles as a result of the discrete sampling of the
mass density field.  We define the error covariance matrix between two
quantities $X$ and $Y$ as $<\!\Delta \! X \, \Delta\! Y\!>$ where the
symbol $<.>$ denotes averaging over many different discrete samplings
of the underlying mass field with the same selection function and
number of particles as the original flux limited distribution. The $\Delta$ denotes the
difference between the true and recovered values. The true value being
obtained by application of LAP on a sampling with an {\it infinite}
number of particles, but with the same selection function as in the
dilute distribution.  We focus on computing the velocity error
covariance matrix.  The calculation for the recovered positions is
similar.  Using (\ref{expandv}), The velocity error covariance matrix
between two particles $i$ and $j$ can be written in terms of the
coefficients $\vC$ error matrix,
\begin{equation}
<\!\Delta \! \valpha_i\,  \Delta \valpha_j\!>=
\sum_{m,n} p_n p_m <\!\Delta \!\vC_{i,n}  \,
\Delta\!\vC_{j,m}\!> \, ,
\end{equation}
By setting to zero the
action gradients (\ref{ceqn:ortho}) we find
\begin{equation}
<\!\Delta \!\vC_{i,n}  \,
\Delta\!\vC_{j,m}\!>=
\frac{9}{4} \int_0^1\!\dd D \dd D^\prime \frac{q_n q^\prime_m }{(D D^\prime)^{1/2}} 
<\!\Delta\!\vg_i\, \Delta\!\vg_j^\prime\!> 
 \, ,
\label{covc}
\end{equation}
were quantities with and without the prime symbol are evaluated at
times $D^\prime$ and $D$, respectively. This expression involves the
gravity error covariance matrix computed between errors at different
times.  We can estimate this matrix under the assumption that the
deviations from the Zel'dovich solution do not 
affect on its value.  According to the Zel'dovich
approximation, the gravity force acting on a particle at a time
$D$ is $\vg_i = D(f^2/\Omega) \vg_{i,0} $, where $\vg_{i,0} $ is computed at 
the final time.  With this approximation (\ref{covc}) becomes
\begin{equation}
<\!\Delta \!\vC_{i,n}  \,
\Delta\!\vC_{j,m}\!>=
B_{n,m} <\!\Delta\!\vg_{i,0}\, \Delta\!\vg_{j,0} \!> \, ,
\label{covca}
\end{equation}
where $B_{n,m}=(9/4) \int_0^1 q_n q^\prime_m
(D{D^\prime})^{1/2} (f f^\prime)^2/(\Omega \Omega^\prime) 
\dd D \dd D^\prime$.  The calculation of the gravity 
error covariance  at the final time is straightforward (Yahil et. al. 1991) 
and the result is
\begin{equation}
<\!\Delta\!\vg_{i,0}\, \Delta\!\vg_{j,0} \!>=
\frac{1}{(4\pi n_1)^{1/2}} \sum_k 
\frac{1}{\phi^2_k}\frac{(\vx_i-\vx_k)(\vx_j-\vx_k)}
{|\vx_i-\vx_k|^3 |\vx_j-\vx_k|^3 } 
\end{equation}

\subsection{Redshift space distributions}

Typically galaxy surveys provide redshifts and angular positions of
galaxies in the sky.  Redshifts of galaxies differ from their
distances as a result of peculiar velocities along the line of sight.
This causes differences between the distribution of
galaxies in real and redshift space (e.g., Kaiser 1987, Hamilton 1993). 
These differences are 
referred to  as redshift distortions.
Method for recovering  the  velocity from the redshift space
distribution of galaxies in the nonlinear regime have so far relied 
on iterations between redshift  and 
real space (Shaya et al 1995, Yahil et. al.
1991, Schmoldt and Saha, 1998). 
Here we show that 
FAM can be extend to treat redshift distortions by
direct minimization of a modified action, without the use 
iterations. For simplicity, we restrict the analysis  
to volume limited surveys.
We define the comoving {\it redshift} coordinate, $\vs_0$, of a galaxy
at the present time, as
\begin{equation}
\vs_0=H_0 \vx_0+ (\vv_0\cdot \hat \vs_0) \hat \vs_0 \, \label{red}
\end{equation}
the $0$ subscript 
refers to quantities at the present time and 
the unit vector  $\hat \vs_0$ points
in the direction of the line of sight to the galaxy. 
The comoving peculiar velocity of the galaxy is  $\vv=(\dd \vx/\dd t)
=D H f(\Omega)  \valpha$. 
 
The redshifts of galaxies are given. So  the expansion of orbits
in terms of time-dependent functions has to be such that the redshift
coordinates as given by (\ref{red}) are fixed under variations of the
expansion coefficients.  Defining parallel ($^\parallel$) and
perpendicular ($^\perp$) directions to the line of sight at the
present time, we write the expansion of the orbits as
\begin{eqnarray}
\xpar_{i}(D)&=&H_0^{-1}\vs_{0,i} +
\sum_n q_n(D) \cpar_{i,n} -f_0\sum_n p_{0,n} \cpar_{i,n} \nonumber \\
&=& \vs_{0,i} + \sum_n {Q}_n(D) \cpar_{i,n} \nonumber \\ 
\xper_{i}(D)&=& \sum_n q_n(D) \cper_{i,n} \, \label{decomp}
\end{eqnarray}
where $p_{0,n}=p_n(D=1)$ and
${Q}_n(D)= q_n(D)-f_0 p_{0,n}$.
This expansion of the orbits satisfies the boundary conditions of 
fixed redshifts and angular positions on the sky.
The role of the Hubble constant $H_0$ is adjusting the units. 
The trivial dependence on $H_0$ can be completely eliminated 
by working with $H_0 \vx$ instead of $\vx$.  
Stationary first variations of the action (\ref{lapD}) with respect to
$\cper$ subject to the boundary conditions (\ref{boundary}) yield
\begin{equation}
\int_0^1 \dd D D^{3/2} {q}_n 
\left[\frac{\dd \vtheta^\perp_i}{\dd D} +
\frac{3}{2}\frac{\vtheta^\perp_i}{D} -
\frac{3}{2}\frac{1}{D^2}\frac{\Omega}{f^2}\vg^\perp_i \right]=0 \,
\label{perp:cor}
\end{equation}
which is 
the time averaged  equation of motion of a particle 
in the plane perpendicular to its  sight-line at
the present time.  On the other hand, stationary variations with
respect to $\cpar$ yield
\begin{equation}
{Q}_{0,n}\valpha^{\parallel}_{0,i}
-\int_0^1 \dd D D^{3/2} {Q}_n 
\left[\frac{\dd \vtheta^\parallel_i}{\dd D} +
\frac{3}{2}\frac{\vtheta^\parallel_i}{D} -
\frac{3}{2}\frac{1}{D^2}\frac{\Omega}{f^2}\vg^\parallel_i \right]=0 \,
\label{wrong}
\end{equation}
where ${Q}_{0,n}={Q}_n(D=1)=-f_0 p_{0,n}$ and 
$\valpha^\parallel_{0,i}=\sum p_{0,n}\cpar$.  These equations differ
from the time averaged equations of motion by a boundary term. 
This term can be eliminated by adding to  the action (\ref{lapD}) 
a kinetic energy term corresponding to the line of sight parallel degree
of freedom, as follows
\begin{equation}
{\cal S} = {\mathrm S}  +
\frac{1}{2} \sum_i \left(\valpha_{0,i}^\parallel\right)^2 \, .
\label{lapDN}
\end{equation}
Minimization of the  the modified action  
readily yields the orbits expansion parameters $\vC_{i,n}$
(see Schmoldt and Saha 1998, for a similar treatment of this problem).

The recovered orbits from redshift space depend on $\Omega$ through
the $f_0$ in the expansion (\ref{decomp}).  The effect of this
dependence in the linear regime is elucidated in Nusser \& Davis
(1994).  Since linear theory overestimates the true velocities (Nusser 
et. al. 1991), this
$\Omega$ dependence will be weaker in the nonlinear regime.
\subsection{Biased distributions}
So far we have assumed that the number of galaxies in a small cell is
proportional the the average mass density in the cell.  However,
galaxies are most likely to be biased tracers of the mass distribution
as indicated by the relative bias between galaxies of different
luminosities and morphological types (e.g, Loveday et al 1995). 
For simplicity  of notations we discuss here how our
scheme can incorporate biasing only for volume limited galaxy
distributions in real space.  Suppose we are given smooth versions of
the galaxy number density and the mass density fields.  Working with
smoothed fields seems reasonable because galaxy formation at a given
point is likely to be affected by the nearby dark matter environment.
We also assume that the smoothing scale is large enough such that the smoothed
galaxy number density field is not contaminated by shot-noise.  Let
$\delta^g$ and $\delta$ be, respectively, the galaxy number density
contrast and the mass density contrast, both smoothed with the same
smoothing window  of width fixed by the physical processes involved in 
galaxy formation.  For unbiased galaxy distribution
$\delta^g=\delta$ and for the familiar linear biasing $\delta^g=b
\delta$ where $b$ is the linear bias factor.  Here 
$\delta^g$ is a nonlinear function of $\delta$, which we assume 
to be local and deterministic (e.g., Dekel \& Lahav 1998). We
characterize the biasing relation at any point in space by the ratio
\[{\cal W}\equiv \frac{1+\delta}{1+\delta^g}.\] Our definition of
biasing in terms of smooth fields inevitably implies that the galaxy
distribution does not contain information on the structure of mass
density on scales smaller than the smoothing scale length.
Given $\cal W$ we
define the following density field
\begin{equation}
 \varrho(\vx) =  \frac {\rho_b}{\bar n} {\cal W}(\vx)
\frac{1}{V} \sum_i\delta^{\rm
D}\left(\vx-\vx_i\right)
 \, . \label{bias}
\end{equation}
This field serves as an unbiased estimate
of smoothed underlying mass distribution. The gravitational force
field, $\vg$, is then estimated from $ \varrho$ by
\begin{equation}
\vg(\vx)=
-\frac{1}{4\pi   {\bar n} }\sum_i {\cal W}_i \frac{\vx-\vx_i} 
{|\vx-\vx_i|^3} 
+\frac{1}{3}\vx , \label{gforce:bias}
\end{equation}
where ${\cal W}_i={\cal W}[\vx_i(D)]$.  If the galaxies and the mass
particles share the same velocity field then the continuity equation
implies that ${\cal W}$ remains constant along the streamlines so that
${\cal W}[\vx_i(D)]= {\cal W}(\vx_{i,0})$ (e.g., Nusser \& Davis 1994,
Fry \& Gaztanaga 1995).  Therefore, it is sufficient to specify the biasing
relation at the present time.  The net effect of biasing is that it
changes the weight assigned to each particle in the calculation of
gravitational fields. This can readily be incorporated in the
TREECODE by assigning to each particle a mass
proportional to $W(\vx_{i,0})$.

\section{discussion}
We have presented  a fast method for  solving the boundary value problem of
recovering the orbits of particles from their present positions
assuming homogeneous initial conditions. The method which we term FAM
is based on P89 implementation of lest action principle
in a cosmological contest. 
It can 
be  applied to  distribution of galaxies in redshift space.
It  can also very easily incorporate any local biasing relation.
FAM  is suitable  for  recovering  orbits from large galaxy
redshift surveys such as the PSC$z$.  It can also be applied to large
portions of the future Sloan Digital Sky Survey.

We have described the method assuming that the number of expansion
coefficients is the same for all galaxies. However, this need not be
the case. For example, galaxies in low density regions can be assumed
to move along straight line just like in the Zel'dovich approximation.
This can significantly speed up FAM especially for the flux limited
surveys with dilute galaxy distribution at large distances from the
observer.

We have used an N-body simulation to show that FAM recovers very well
the final velocities from a given volume limited particle distribution
in real space.  However,
galaxy surveys provide galaxy distribution in redshift space.  Redshift
distortions introduce the nuisance of multi-valued zones where
particles overlap in redshift space while they are far apart in real
space.  FAM allows a recovery of the orbits non-iteratively from redshift space
data by direct minimization of a modified action. We believe that this should 
mitigate the effects of multi-values zones in the recovered orbits.
Tests of FAM recovery from both flux and volume limited 
distributions in redshift space are underway. 

In this work we concentrated on how well FAM can reconstruct
objects' peculiar velocities and we did not
check  how well it can recover initial
fluctuations. Judged by its superiority over the Zel'dovich solution
at recovering the present velocities, FAM is expected to perform well
at recovering the density fluctuation at any time in the past. We have
outlined FAM can incorporate possible biasing between the mass and
galaxies. We have shown that we only need to specify a biasing relation
at the present time.  The recovery of the orbits is sensitive to the
biasing relation.  So one can  tune  the biasing
relation so  that the clustering amplitude
of the recovered mass
density field varies with time according to
hierarchical structure formation. 
In hierarchical clustering, the evolution of 
clustering amplitude, as measured for example by the 
correlation function, is almost independent of the linear mass power spectrum
(e.g., Hamilton et. al. 1991, Jain,
Mo \& White 1995, Peacock \& Dodds  1995), and of the cosmological parameters
if expressed as a function of the linear density growing mode, $D$ 
(e.g., Nusser \& Colberg 1997). Therefore, one can determine the biasing 
relation independent of the cosmological parameters and the  details of the 
dark matter model.

\section{acknowledgment}
AN is grateful to the Astronomy
Department at UC Berkeley for its support of a visit during which this
work has been completed.
EB thanks the MPA of Garching, the Hebrew University of Jerusalem and
the Technion of Haifa for their hospitality  while part of this work was done.
AN has benefited, over the course of several years,
from many  stimulating  discussions
with Marc Davis and Simon White on gravitational dynamics.
We wish to thank Shaun Cole for allowing the use of the N-body simulation.

\end{document}